\def\spose#1{\hbox to 0pt{#1\hss}}
\def\lta{\mathrel{\spose{\lower 3pt\hbox{$\mathchar"218$}}
     \raise 2.0pt\hbox{$\mathchar"13C$}}}
\def\gta{\mathrel{\spose{\lower 3pt\hbox{$\mathchar"218$}}
     \raise 2.0pt\hbox{$\mathchar"13E$}}}
\documentstyle[aas2pp4]{article}

\slugcomment{Submitted to The Astrophysical Journal (Letters)}

\lefthead{Walker and Wardle}
\righthead{ESEs and dark matter}

\begin{document}

\title{Extreme scattering events and Galactic dark matter}

\author{Mark Walker and Mark Wardle}
\affil{Special Research Centre for Theoretical Astrophysics,}
\affil{School of Physics, University of Sydney, NSW 2006, Australia}

\begin{abstract}
``Extreme Scattering Events'' (ESEs) are attributed to
radio-wave refraction by a cloud of free-electrons crossing the
line-of-sight. We present a new model in which these electrons form
the photo-ionized `skin' of an underlying cool, self-gravitating
cloud in the Galactic halo. In this way
we avoid the severe over-pressure problem which afflicts other
models. The UV flux in the Galactic halo naturally generates
electron densities of the right order. We demonstrate, for the first
time, a good reproduction of the prototypical ESE in the quasar 0954+658.
The neutral clouds are a few AU in radius and have masses $\lta10^{-3}\,
{\rm M_\odot}$. The observed rate of ESEs implies that a large fraction
of the mass of the Galaxy is in this form. 
\end{abstract}


\section{Introduction}
ESEs were discovered ten years ago during radio flux monitoring
of a sample of compact radio quasars (Fiedler et al 1987).
The ESE phenomenon consists of dramatic flux changes
occuring over several weeks to months.
It is broadly agreed that ESEs are not intrinsic
variations but, rather, apparent flux changes which are
caused by refracting elements, a few AU in radius,
crossing the line-of-sight.
Both random (Fiedler et al 1987) and deterministic
(Romani, Blandford \& Cordes 1987) lens structures have been
proposed; in all cases the refraction is attributed to free
electrons. Two further points of consensus are that the blobs
of free electrons must be Galactic,
and that they represent a distinct component of the ISM
(Narayan 1988; see also Rickett 1990). This same component is
thought to be responsible for episodes of multiple
imaging of radio pulsars (Cordes \& Wolszczan 1986;
Rickett 1990) --- a phenomenon which manifests itself as
periodic fringes in the dynamic spectra.

Despite the attention which the ESE phenomenon has attracted
(Romani, Blandford \& Cordes 1987; Romani 1988; Clegg,
Chernoff \& Cordes 1988; Clegg, Fey \& Lazio 1998),
the current state of understanding is unsatisfactory for
two reasons. First the implied pressure of the electron cloud
-- having an inferred density $n_e\sim10^3\;{\rm cm^{-3}}$
and a temperature ${\rm T\sim10^4\;K}$ 
-- exceeds that of the diffuse Interstellar Medium (ISM) by
a factor of $10^3$. Such a cloud should dissipate on a time-scale
comparable to the sound-crossing time, of order a year.
This difficulty can be lessened by appealing
to cloud geometries which are elongated along the line-of-sight,
and by placing the clouds in regions of high pressure within
the ISM --- specifically, old supernova shocks. In support of
the latter idea it has been argued (Romani 1988) that the
distribution on the sky of recorded ESEs reflects that of
the strongest regions of non-thermal radio emission. However,
with the small number of ESEs recorded to date, and the large
area covered by the radio loops, this evidence is at best
suggestive. The second main difficulty with existing models
is that none has been able to reproduce the dual-frequency
light-curve of the quasar 0954+658 --- the source for which
the best ESE data exist.

In this paper we present the essential elements of
a new model for ESEs, wherein the free electrons
comprise a `skin' of photo-ionized material around
a cool, self-gravitating cloud. The pressure of the
neutral material is balanced by the self-gravity of the cloud,
while the ionized gas flows continuously from the surface;
complete evaporation occurs over a time-scale of order
the Hubble time. We show in \S2 that this model reproduces
the prototypical ESE light-curves. The event rate for ESEs
then leads to the robust conclusion (\S3) that these clouds contribute
substantially to the Galactic dark matter.
In \S4 we note some key observational tests.

\section{Refraction in a photo-evaporated wind}
The estimated cloud dimension and distance indicate that
geometric optics is a good approximation. To proceed we
need to know the electron density distribution, from which
we can calculate the bending angle of rays, and hence the
imaging properties of the cloud.

The photo-ionized material at the surface of the cloud is
not bound, but flows away as a feeble wind. Dyson (1968)
constructed an analytic solution for the density profile
of a spherically symmetric photo-evaporated wind from a
neutral cloud; his solution is plotted in figure 1. The
electron density at the surface of the cloud is given
(in ${\rm cm^{-3}}$) by $4.87\times10^6\;\sqrt{J_\infty/R}$,
and in the local Galactic halo, at the solar circle, the
ionising photon field is $J_\infty\simeq10^6\;
{\rm cm^{-2}\,s^{-1}\,sr^{-1}}$ (Dove and Shull 1994).
For a cloud radius $R\simeq3\times10^{13}\;{\rm cm}$ (Fiedler
et al 1987) we obtain $n_e(R)\simeq10^3\;{\rm cm^{-3}}$. This
model thus exhibits immediate strength in that the electron
density required to explain the ESEs is a {\it prediction\/}
of our model, whereas it is an {\it assumption\/} of existing
models.

Dyson's solution cannot be used directly for our calculations,
however, as it is discontinuous at the ionisation front,
whereas we require the first and second derivatives of $n_e$
to be well behaved. Physically the thickness of the ionisation front
is just the absorption length of Lyman-limit radiation in the
neutral medium. In turn this is fixed by the temperature of the
neutral atmosphere, via the requirement for pressure balance
across the transition region. Another difficulty with Dyson's solution
is that $n_e$ is determined implicitly, which is computationally
inconvenient. We therefore employ the approximation
$n_e(x)\propto F(x)\,x^{-2} \,(1+4\log_ex)^{-1/2}$,
where $x$ is the radius expressed in units of the cloud radius,
and $F$ is the Fermi function, taken to have a width of
$\delta x=0.01$ (corresponding to a temperature of 40~K in the
neutral atmosphere). This distribution is plotted in figure 1.

\placefigure{fig1}

Making use of this electron density distribution,
together with the phase-screen approximation,
it is straightforward to compute light-curves
for a neutral cloud traversing the line-of-sight to a distant
radio source. Figure 2 shows one such computation, appropriate
to a cloud at a distance of 2~kpc and an impact parameter of $R/2$,
along with the data for 0954+658. To simulate the non-zero size
of the real source, the theoretical curves are smoothed by
convolution with Gaussian functions. The adopted source model corresponds
to 50\% of the source flux being contained within a compact (lensed)
component, with this component having a brightness temperature
of $8\times10^{11}$~K.

\placefigure{fig2}

The main qualitative features of these light curves are
accounted for as follows. The phase velocity of the wave is increased
by the presence of free electrons, so the cloud acts as a diverging
lens. At low frequencies the lens is powerful enough that almost all
rays are refracted out of the line-of-sight, and only a small flux is
measured when the lens is aligned with the source; this behavior is
generic to all blobs of free electrons regardless of the details of
their density distribution. Consequently this regime of a very strong
lens is not particularly helpful in distinguishing our model from other
possible electron density distributions.
At higher frequencies, however, the refractive
index of the lens is much smaller, and typically rays are no-longer
refracted through sufficiently large angles that caustics form. Exceptions
occur for rays that pass near the edge of the cloud: here
the photoionized skin creates a large phase curvature in the wavefront
and can introduce caustics even at high frequencies. These caustics are
evident as sharp peaks in the model light-curve, and we note that similar
features appear in the data; in our model they are intimately associated
with the presence of a peak in the electron column density at
the limb of the cloud.  The good qualitative agreement between our
model and the data suggests that surface photo-evaporation
-- which, of course, implies underlying {\it cool\/} material
-- is an essential feature of real ESE clouds. By contrast, the
Gaussian electron density profile originally proposed (Romani, Blandford
and Cordes 1987) for ESEs cannot, even qualitatively, match the dual
frequency light curve of 0954+658.

\section{Implications for dark matter}
Going beyond the interpretation of individual events, the
principal implication of the new model is that there is
much more mass present in the ESE clouds than had been
previously thought (cf. Fiedler et al 1987; but see also
Pfenniger \& Combes 1994). We now derive
a lower limit on the total contribution of the ESE
clouds to the mass of the Galaxy.

The sky-covering factor, $f$, of the clouds is estimated
from the flux monitoring data (Fiedler et al 1994) to be
$f\sim5\times10^{-3}$.
We can immediately relate $f$ to the total surface density,
$\Sigma$, in clouds at the solar circle: $\Sigma\simeq
2\,\langle\sin|{\rm b}|\rangle f\,M/\pi R^2\sim f\,M/\pi R^2$,
for cloud mass $M$ and radius $R$ (ESEs do not preferentially
occur at low Galactic latitude, b, so we have set
$\langle\sin|{\rm b}|\rangle\sim0.5$). These quantities are in
turn related by the requirement of hydrostatic equilibrium
within each cloud -- $k{\rm T}\sim GMm_p/R$ at temperature T
-- leading to $\Sigma\sim fk{\rm T}/\pi Gm_pR$. The cloud radius
can be inferred from the event duration,
in combination with an assumed transverse speed; a limit
follows from setting the transverse speed equal to the
escape speed for the Galaxy ($500\;{\rm km\,s^{-1}}$), giving
$R\lta3\times10^{14}\;{\rm cm}$. We expect that the cloud temperatures
are at least as large as that of the cosmic microwave background
(i.e. ${\rm T}\gta3$~K), so we deduce
\begin{equation}
\Sigma\;\gta\;10^2\;\,{\rm M_\odot\;pc^{-2}}.
\end{equation}
This lower limit is already
larger than the total surface density of observable matter,
and consistent with the value ($210\;{\rm M_\odot\;pc^{-2}}$)
necessary to explain the Galactic rotation curve (Binney
and Tremaine 1987). Taking the dynamically determined surface density
as an upper limit on the contribution from ESE clouds, we
estimate the individual cloud masses to be
$\lta10^{-3}\,{\rm M_\odot}$: highly sub-stellar. Hence
the ESE data lead us to the conclusion that low-mass gas
clouds make up a good fraction of the mass of the Galaxy.

We are currently reassessing the sky covering fraction, $f$
(manuscript in preparation), based on our theoretical model
and the data of Fiedler et~al. (1994). It is possible that $f$
is as small as a few times $10^{-4}$, in which case the
lower limit on $\Sigma$ would be an order of magnitude
smaller. Even at this level, however, the cool, low-mass cloud
population implied by the model remains a substantial and
poorly understood component of the Galaxy. Moreover, because
our estimate is a lower limit on the surface density, a smaller
covering fraction still admits the possibility that such
clouds do indeed dominate the dynamics of the Galaxy.

\section{Discussion}
The result we have presented might appear surprising; can dark
matter really be in this form and yet have escaped detection?
It has been noted previously (e.g. Pfenniger, Combes \& Martinet
1994, Gerhard \& Silk 1996) that cool, dense gas clouds are not easy
to detect directly and may indeed be a viable dark matter
candidate. We reconsider these issues elsewhere (manuscript in
preparation), but some brief comments are appropriate here.

First, by virtue of their high density ($\sim10^{12}\;{\rm cm^{-3}}$)
and low temperature, three body reactions are sufficient to ensure
that the clouds are molecular. The constituent material
must have a very low dust/gas ratio in comparison with the
diffuse ISM, otherwise the population would already have
been revealed by optical extinction events of extragalactic
stars (notably in the data generated by the microlensing
searches: Paczy\'nski 1996). A low dust content might reflect
either a very low metal abundance or, possibly
(B. Paczy\'nski, personal communication),
settling of the refractory elements into a small, rocky core.
Thirdly, unless the clouds are dynamically unimportant (i.e.
$\Sigma\ll10^2\;\,{\rm M_\odot\;pc^{-2}}$), the measured
$\gamma$-ray background at high Galactic latitudes
(Kniffen et~al. 1996) constrains the vertical distribution
of the clouds to be much more extensive than the cosmic
ray disk, implying that the clouds are a halo population.
(In turn this means that they are pre-Galactic objects.)
The small covering fraction, $f$, of the
clouds means that each can make $1/f$ orbits
of the Galaxy before colliding with another cloud, implying
a lifetime of order the Hubble time. The large column density
of the individual clouds (of order $10^{25}\;{\rm cm^{-2}}$)
means that interactions with the diffuse gas in the Galactic
plane involve a negligible deceleration. The lifetime against
UV photo-evaporation is also of order the Hubble time. 

While the cool-cloud model we advance for
ESEs is compatible with a variety of existing data, some
specific observations can be made which would provide strong tests
of the picture. We offer three such tests; the first of
these is most intimately connected with ESEs, and is in some
sense the easiest experiment to perform. Although the clouds are
predominantly molecular, a crude assessment of the relevant chemistry
suggests that they should contain sufficient atomic hydrogen
to make them opaque in the 21~cm line. Thus, while an ESE
is in progress there should also be a narrow absorption line
present, in addition to the normal absorption spectrum of the source.
(In our model, a similar connection should also hold between
21~cm absorption and multiple imaging of pulsars.)
Small-scale (tens of AU) structure has already been reported
in 21~cm absorption lines (Davis, Diamond \& Goss 1996, Frail et al 1994),
but the possible connection with ESEs has not yet been explored.
The line velocities should reflect the kinematics of the cloud
population, which are expected to be quite different to that
of the diffuse HI in the Galactic disk. We note that, because of
the multiple imaging which takes place, the absorption lines
may appear saturated but with a non-zero flux minimum.

Secondly, because the inferred column density of neutral material is
so large it is expected that the
clouds are opaque to Rayleigh scattering (c.f. Combes \&
Pfenniger 1997) between $\lambda\simeq2000\,{\rm\AA}$ and the
Lyman/Werner bands of ${\rm H_2}$; and opaque to Thomson scattering at
X-ray energies. Thus one in $1/f$ extragalactic UV/X-ray sources (of
sub-milliarcsecond size) should appear extinguished at any one time,
with events lasting a month or so.

Finally we note that the
recombination lines (H$\alpha$ etc) from the cloud surface may
be detectable if there is a sufficiently strong local source
of ionising radiation (Gerhard \& Silk 1996). That is, clouds
which are within a parsec of a hot star will be luminous sources
of H$\alpha$ and may thus be rendered visible. It might also be
possible to detect, from their H$\alpha$ emission, the clouds
which are closest to the Sun. Many clouds are expected within a
parsec, and these may appear as compact H$\alpha$
nebulae of large radial velocity and very high proper-motion.

\section{Conclusions}
We have presented a new model for the cause of Extreme Scattering
Events, where radio-wave refraction by ionized material magnifies
(``lenses'') a distant radio source. In our model the ionized
material is generated by photo-evaporation of an underlying neutral,
self-gravitating cloud, so there is no difficulty in understanding the
requisite high electron pressures; indeed the ambient UV field in the
Galactic halo naturally generates the necessary electron density.
Moreover we find it straightforward to reproduce the dual-frequency
light-curve of the prototypical ESE in the source 0954+658. These facts
argue strongly for the soundness of the model.

It follows immediately that a large fraction of the mass of the
Galaxy is in the form of cool gas clouds of up to Jovian mass,
and only a few AU in radius. We are unable to falsify this picture
with existing data, but with new observations some straightforward
tests can be made.

\acknowledgments
Particular thanks go to Ron Ekers for many helpful discussions; we
are also grateful to Simon Johnston, B\"arbel Koribalski, Don Melrose,
Bohdan Paczy\'nski, Gordon Robertson and Lister Staveley-Smith for their input.



\plotone{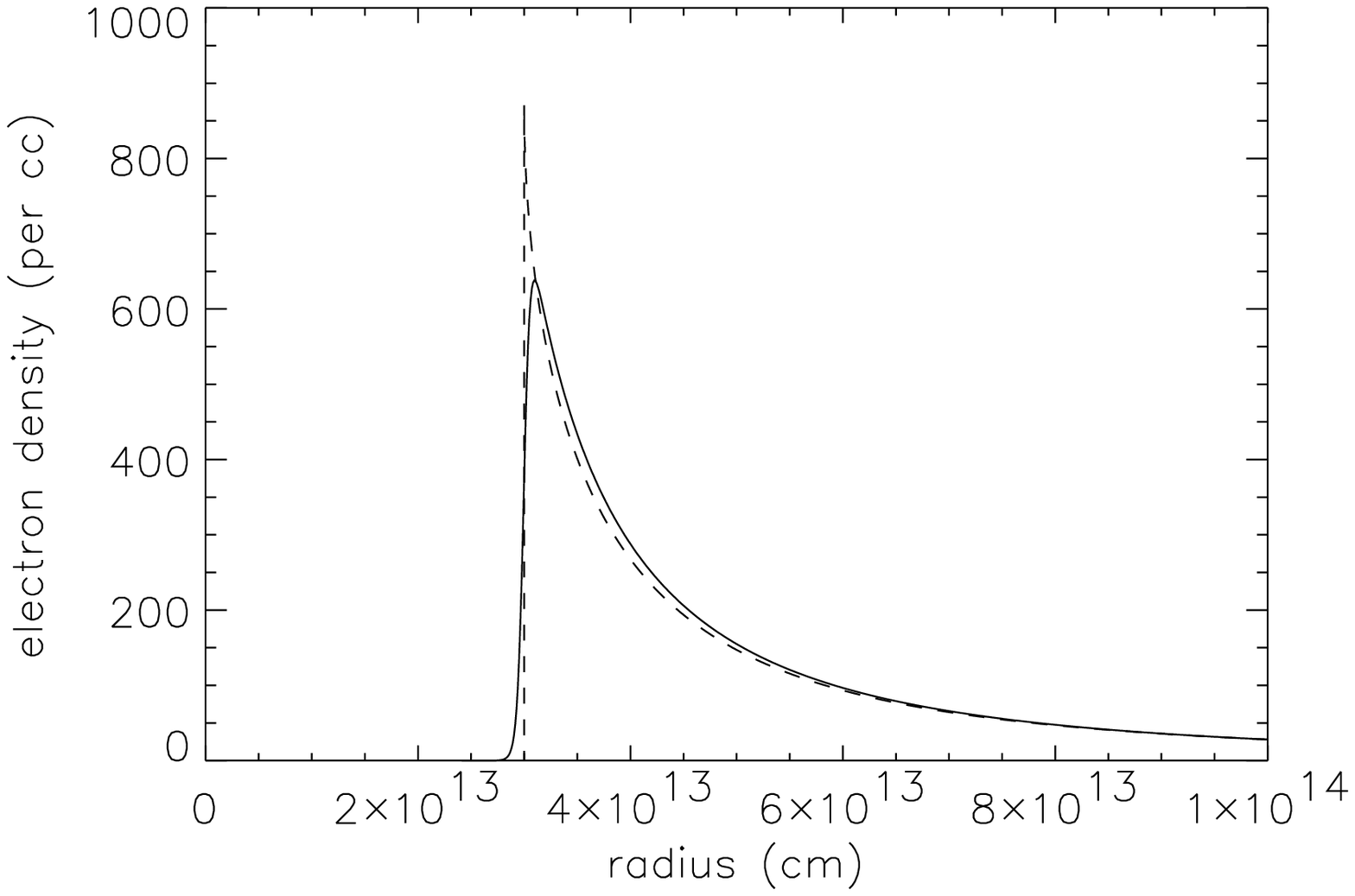}
\figcaption{Electron density as a function
of radius for a photo-evaporated wind: Dyson's (1968) solution
(dashed line); and our approximation (full line).\label{fig1}}

\clearpage
\plotone{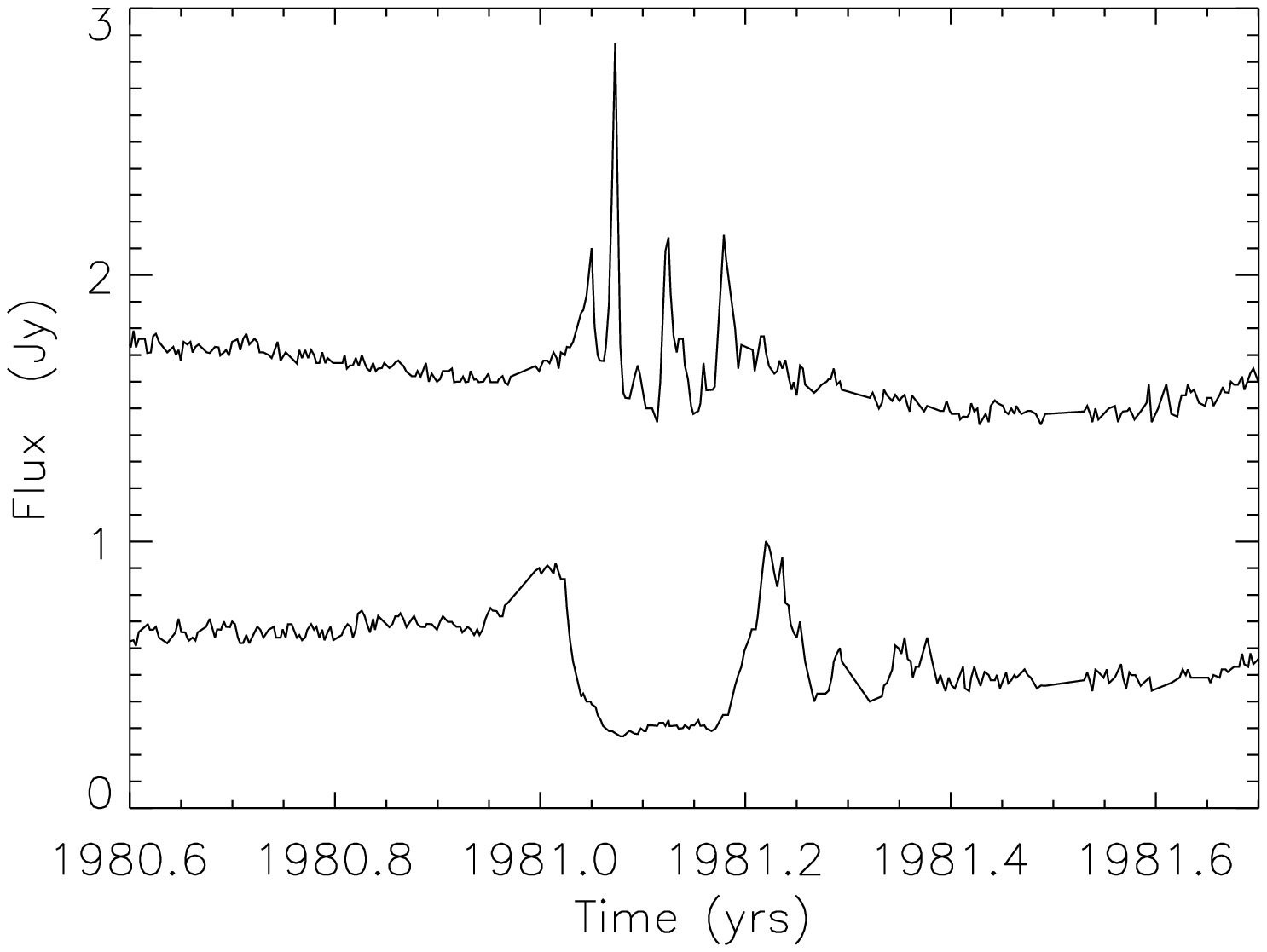}

\plotone{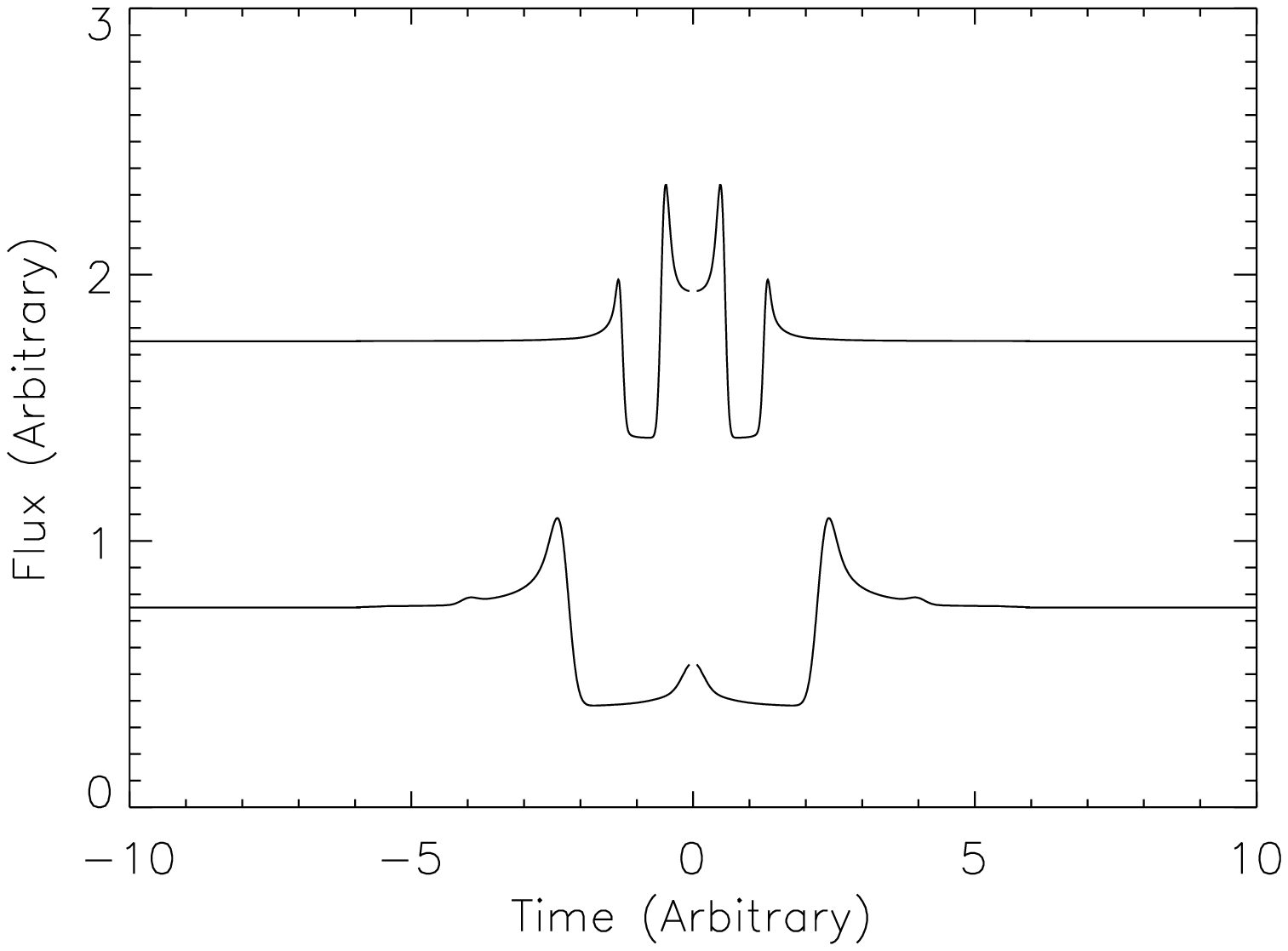}
\figcaption{Observations (top panel, adapted from
Clegg, Fey and Lazio [1998]; original data in Fiedler et al. [1987])
and theory (lower panel) for the ESE seen in the quasar 0954+658.
The upper curves correspond to a radio frequency of 8.1~GHz (to which
1~Jy has been added), and the lower curves to 2.7~GHz.\label{fig2}}

\end{document}